\documentclass[journal,12pt,draftclsnofoot,onecolumn]{IEEEtran}
\ifCLASSINFOpdf
   \usepackage[pdftex]{graphicx}
\else
\fi
\usepackage[cmex10]{amsmath}
\usepackage[tight,footnotesize]{subfigure}
\usepackage{setspace}
\usepackage{amsfonts}
\usepackage{graphics} % for pdf, bitmapped graphics files
\usepackage{setspace}
\usepackage{amssymb}
\usepackage{epsfig} % for postscript graphics files
\usepackage{stfloats}
\begin{document}
%
% paper title
% can use linebreaks \\ within to get better formatting as desired
%\title{Task-oriented Time Varying Graph Modeling for Space-based Information Network Topology}
\title{Complex Network Theoretical Analysis on Information Dissemination over Vehicular Networks}
% author names and affiliations
% use a multiple column layout for up to three different
% affiliations
%\author{\authorblockN{AA, BB, CC}}

\author{\authorblockN{Jingjing~Wang\authorrefmark{1}, Chunxiao~Jiang\authorrefmark{1}, Longxiang~Gao\authorrefmark{2}, Shui~Yu\authorrefmark{2}, Zhu~Han\authorrefmark{3}, and Yong Ren\authorrefmark{1}} %\vspace{-2mm}
\small\authorblockA{\authorrefmark{1}Department of Electronic Engineering, Tsinghua University, Beijing, 100084, China\\
\authorrefmark{2}School of Information Technology, Deakin University, Burwood, VIC 3125, Australia\\
\authorrefmark{3}Electrical and Computer Engineering Department, University of Houston, Houston, TX, USA\\%\vspace{-2mm}
E-mail: chinaeephd@gmail.com, \{jchx, reny\}@tsinghua.edu.cn,} \{longxiang.gao, syu\}@deakin.edu.au, zhan2@uh.edu}

\maketitle

\begin{abstract}
How to enhance the communication efficiency and quality on vehicular networks is one critical important issue. While with the larger and larger scale of vehicular networks in dense cities, the real-world datasets show that the vehicular networks essentially belong to the complex network model. Meanwhile, the extensive research on complex networks has shown that the complex network theory can both provide an accurate network illustration model and further make great contributions to the network design, optimization and management. In this paper, we start with analyzing characteristics of a taxi GPS dataset and then establishing the vehicular-to-infrastructure, vehicle-to-vehicle and the hybrid communication model, respectively. Moreover, we propose a clustering algorithm for station selection, a traffic allocation optimization model and an information source selection model based on the communication performances and complex network theory.
\end{abstract}

% For peer review papers, you can put extra information on the cover
% page as needed:
% \ifCLASSOPTIONpeerreview
% \begin{center} \bfseries EDICS Category: 3-BBND \end{center}
% \fi
%
% For peerreview papers, this IEEEtran command inserts a page break and
% creates the second title. It will be ignored for other modes.
\IEEEpeerreviewmaketitle

\section{Introduction}
% no \IEEEPARstart
\label{Introduction}
Due to the emerging of intelligent transport system, vehicular networks have received lots of attentions. Although cellular networks enable convenient voice communication and simple entertainment services to drivers and passengers, they are not well-suited for certain direct vehicle-to-vehicle (V2V) or vehicle-to-infrastructure (V2I) communications~\cite{hartenstein2008tutorial}. In particular, how to improve the performances of the communication system has already been under development~\cite{giust2015distributed}, where some key technologies~\cite{demestichas20135g}, e.g., small cells, device-to-device (D2D) communication, mobile clouds, flexible spectrum management etc., can be considered to be employed in vehicular networks.

In the literature of vehicular networks, many researches focused on improvement of the vehicle mobility models~\cite{sharef2014vehicular}, communication channel models and the routing strategies~\cite{zhang2015modeling}~\cite{jiang2014data}, while the network properties as well as the complex characteristics of the vehicular networks have not been fully investigated. The vehicular networks are associated with a tremendous network size. Moreover, diverse hierarchical structures and node types give rise to more complex interactions. Furthermore, vehicular networks have a complex time-space relationship. The mobility of the vehicles on the road lead to the dynamic evolutionary topology. In terms of some hot communication technologies, the ultra dense cellular deployment would lead to more than ever interactions among vehicle units (vehicles to infrastructures and infrastructure to infrastructure) and the D2D based vehicular-to-vehicular communication also lead to a more complex hybrid communication network. Therefore, it is necessary to view the vehicular networks from the other dimension, i.e., using complex network theory to discover the complex characteristics of vehicular networks, based on which the network performance can be improved.

With the development of random graph model, the complex network theory emerged based on the~\cite{watts1998collective} and~\cite{barabasi1999emergence}, which discovered the small-word property and the power-law distribution of the node degree of the realistic complex networks. Based on the advantages of complex networks theory, this paper proposes a complex network theoretic view on the vehicular networks with following original contributions.
For one thing, this is the first work to establish the vehicular network V2V and V2I models with complex network theory. Moreover, We use the node degree, average path length, clustering coefficient and betweenness centrality to analyze the topology of a vehicular network based on the taxis GPS database of Beijing~\cite{yuan2010t} and study the relationship between the network topological properties and communication parameters. For another thing, we propose a clustering algorithm, a traffic allocation model and an information source selection model depending on the communication impedance.

The rest of this paper is organized as follows. Section~\ref{Data-Driven complex network model} establishes a vehicular network system model based on the complex network theory, and gives some key parameters and their characters. Section~\ref{The 5G Communication on the VANET} describes three typical vehicular communication models and three optimization algorithm models. Section~\ref{Simulation Results} gives the simulation results for the proposed models. Concluding remarks and future work are given in Section~\ref{Conculsion}.

%%%%%%%%%%%%%%%%%%%%%%%%%%%%%%%%%%%%%%%%%%%%%%%%%%%%%%%%%%%%%%%%%%%%%%%%%
%%%%%%%%%%%%%%%%%%%%%%%%%%%%%%%%%%%%%%%%%%%%%%%%%%%%%%%%%%%%%%%%%%%%%%%%%\section{Related Works}
%\begin{figure*}[!t]
%
%\begin{center}
%  \centering
%\subfigure[Vehicles space distribution with statistical radius $r=200\mathrm{m}$]{\includegraphics[width=0.23\textwidth]{sp1.eps}}
%  \hspace{0.05in}
%
%\subfigure[Vehicles space distribution with statistical radius $r=400\mathrm{m}$]{\includegraphics[width=0.23\textwidth]{sp2.eps}}\\
%  \hspace{0.05in}
%
%\subfigure[Vehicles space distribution with statistical radius $r=600\mathrm{m}$]{\includegraphics[width=0.23\textwidth]{sp3.eps}}
%\hspace{0.05in}
%
%\subfigure[Vehicles space distribution with statistical radius $r=800\mathrm{m}$]{\includegraphics[width=0.23\textwidth]{sp4.eps}}
%\caption{Vehicles space distribution with different specified statistical radius.}
%\label{sp}
%\end{center}
%\end{figure*}
\section{Data-Driven Complex Network Model}
\label{Data-Driven complex network model}

\subsection{Dataset Analysis}
\label{3-1}
In vehicular networks, vehicles can communicate with each other (V2V), and can also establish communication with the roadside infrastructures (V2I). In this subsection, we construct the complex network model for the vehicular networks based on a real-world dataset, which contains the taxi GPS data of Beijing (longitude from 116.25 to 116.55, and latitude from 39.8 to 40.05) obtained from the Microsoft Research Asia~\cite{yuan2010t}.
%First of all, we analyze the numerical characteristics of the GPS data and give an intuitional understanding of this vehicular network (Section \ref{3-1}). Secondly, based on the 5G communication technologies, both weighted and undirected graph models are established and the concept of communication impedance is proposed (Section \ref{3-2}). Last but not the least, in term of the constructed vehicular network model some key parameters are calculated. A variety of properties shows that a vehicular network is indeed a complex network and the complex network theory provides a new perspective to the network optimization and network management (Section \ref{3-3}).

\begin{figure}[!t]
  \centering
\includegraphics[width=0.46\textwidth]{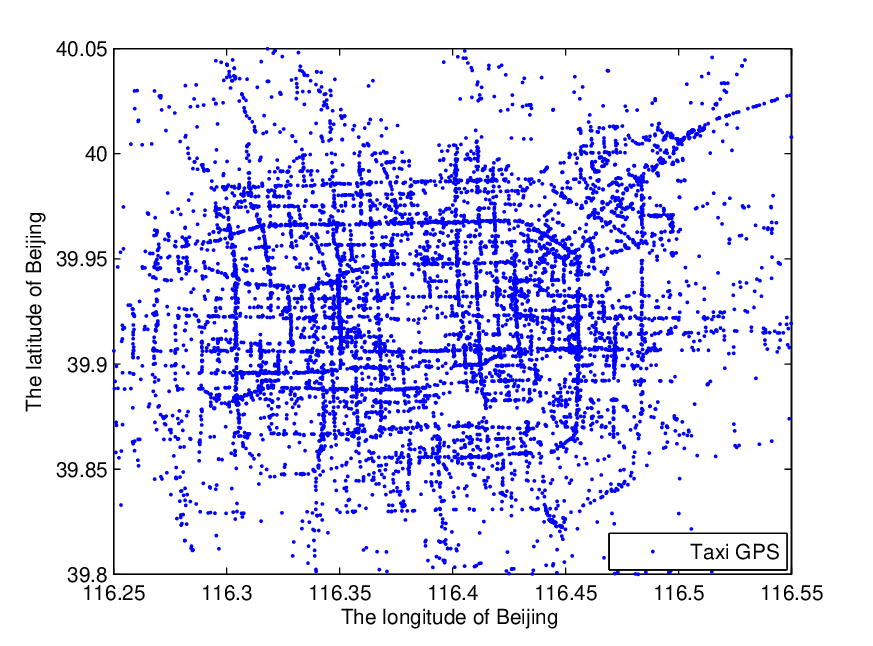}
  \caption{\!The taxis GPS distribution in Beijing (longitude from 116.25 to 116.55 and latitude from 39.8 to 40.05).}\label{taxi GPS}
\end{figure}

Based on the aforementioned GPS dataset, we plot the vehicles position distribution in the Fig.~\ref{taxi GPS} at one moment. The vehicles position distribution clearly reflects the shape planning structure of Beijing and distinguishes its downtown and suburban areas. In the following subsection, we will construct a weighted and undirected graph model based on some key communication parameters for vehicular networks.

%Fig.\ref{sp} shows the space distribution of the vehicles with different statistical radius. The space distribution is a statistic of the vehicle quantities in the areas with specific size. In the random network or the deterministic network, the number of nodes in specific regions follows the Poisson distribution or the uniform distribution. However, the four subgraphs in the Fig.\ref{sp} demonstrate a property of scaling-free in different scenarios. In other words, the space distribution processes a long-tailed property and the traditional statistics in probability theory disable to characterize the network and the traditional research views of vehicular networks based on random networks lack of accuracy and scientific sufficiency.
%In the following subsection, we will construct a weighted and undirected graph model for 5G communication among vehicular networks.

%\begin{figure}[!t]
%  \begin{minipage}[t]{0.46\linewidth}
%  \centering
%  \includegraphics[width=1\textwidth]{sp1.eps}
%  \centerline{\footnotesize{(a) Statistical Radius $r=200\mathrm{m}$}}
%  \end{minipage}
%  \begin{minipage}[t]{0.46\linewidth}
%  \centering
%  \includegraphics[width=1\textwidth]{sp2.eps}
%  \centerline{\footnotesize{(b) Statistical Radius $r=400\mathrm{m}$}}
%  \end{minipage} \\
%  \begin{minipage}[t]{0.46\linewidth}
%  \centering
%  \includegraphics[width=1\textwidth]{sp3.eps}
%  \centerline{\footnotesize{(c) Statistical Radius $r=600\mathrm{m}$}}
%  \end{minipage}
%  \begin{minipage}[t]{0.46\linewidth}
%  \centering
%  \includegraphics[width=1\textwidth]{sp4.eps}
%  \centerline{\footnotesize{(d) Statistical Radius $r=800\mathrm{m}$}}
%  \end{minipage}
%\caption{Vehicles Space Distribution with Different Specified Statistical Radius.}\label{sp}
%%\vspace{2.0cm}
%\end{figure}

\subsection{Weighted and Undirected Graph Models for Vehicular Networks}
\label{3-2}
In accordance with the analyses above, we build the vehicular network model as a weighed and undirected complex network in which the nodes represent the vehicles in the road segments and the undirected edges represent the interaction between the nodes. The interaction in this paper means the communication between each two vehicles. The edges weights measure the communication performances on the vehicular networks which depend on the distance between the communication pairs, communication channel fading, the environment disturbance and the cellular radius.

To simplify modeling and calculation, we assume that the communication ability of each vehicle is identical and communication channel meets the COST 231-Bertoni-Ikegami model~\cite{correia2009view}. In addition, we neglect the cellular gaps and the cellular shapes, which are not affected by terrain. Accordingly, the weighted and undirected vehicular network is noted as a graph $G\!=\!(V,E,R)$, where $V$ is the set of vertices representing vehicles and $E$ is the set of edges representing the interaction among the vertices. Weights $R$ reflect the communication performance on the vehicular network. $R$ reflects the communication performance in the vehicular ad hoc network, where the specific definition of the communication impedance $R$ is based on the following key communication technologies:
%\begin{table}[!t]
%\caption{List of Key Notations}
%\label{Table.1}
%\begin{tabular}{c|c}
%  % after \\: \hline or \cline{col1-col2} \cline{col3-col4} ...
%  \hline
%  \hline
%Notation & Paraphrasing \\
%  \hline
%$r$ (km) & Specified max communication distance\\
%  \hline
%$G=(V,E,R)$ & Graph expression for vehicle network \\
%  \hline
% $d$ (km) & Actual transmission distance\\
% \hline
%$f_{c} (MHz)$ & Carrier frequency\\
% \hline
%$L_{u} (dB)$ & Path loss in the urban area  \\
%\hline
%$L_{0}$    &   Path loss in the free space \\
%\hline
%$\mathfrak{e}\doteq ENR$    & Average signal energy noise ratio\\
% \hline
% $R_{ij}$  & Communication impedance of $i$ and $j$ \\
% \hline
% $C_{i}$   &     Clustering coefficient of vehicle $i$   \\
% \hline
% $l$       &   Network average path length       \\
% \hline
% $B_{i}$  & Betweenness centrality of vehicle $i$ \\
% \hline
%$r_{c}$  &    Radius of dense handover cellular \\
%  \hline
%  $n_{s}$    & Frequency of cellular switching\\
% \hline
% \hline
%\end{tabular}
%\end{table}

\emph{Channel Model}: Because of the city dotted with tall buildings and luxuriant trees, signals from sources may be attenuated severely to destinations. This paper use the COST 231-Bertoni-Ikegami Model to analyze the transmission path loss. We assume that there exists a line-of-sight transmission path between each two communication-capable vehicles. Therefore, the relatively accurate path loss in the urban area, $L_{u}$ can be calculated as:
\begin{equation}\label{2}
{{L}_{u}}=42.6+26\log d+20\log {{f}_{c}}~~\mathrm{dB},
\end{equation}
where $d$ is the transmission distance and $f_{c}$ represents the signal carrier frequency.

\emph{Ultra Dense Cellular Handover}: Communication system tends to construct a multi-layer heterogeneous network covering base stations and low power micro-stations. In order to improve spectrum efficiency and the transmission capacity, we have made unremitting endeavor on the enhancement of the modulation and encoding methods, while the decrease of cell radius can also result in a sharp increase of system capacity. Therefore, an appropriate communication cell radius improves spatial multiplex ratio and reduces the system power consumption.
Nonetheless, an ultra dense cellular handover means a frequency conversion, more shared-spectrum interferences and more difficulties in multi-point coordination. Spontaneously, the time-delay and handoff dropping probability are both increased due to the ultra dense cellular handover, which increases the impedance of communication of each communication link. We make a statistical calculation of the number of cellular switching on each communication link, noted as $n_{s}$.
Based on the communication channel model and ultra dense cellular handover mentioned above and considering the node degrees and betweenness centralities in the complex network theory, we define the weight of the edge connecting node $i$ and node $j$, marked as $R_{ij}$, which is named as link communication impedance:
\begin{equation}\label{15}
{{R}_{ij}}\!=\!\left\{ \begin{matrix}
   \alpha {{({{k}_{i}}{{B}_{i}}\!+\!{{k}_{j}{{B}_{j}}})}^{\upsilon}}\!+\!\beta {{{{{L}_{u}}}}^{\psi }}\!-\!\mu {{(\vartheta/d_{ij})}^{\xi }}+\zeta{n_{s}} ,~~d_{ij}\leq r  \\
   \infty ,~~ d_{ij}\leq r \\
\end{matrix} \right.
\end{equation}
where $k_{i}$ represents the degree of the node $i$ and $B_{i}$ notes the betweenness centrality of vehicle $i$. $\vartheta$ shows the energy noise ratio, $\alpha, \beta, \mu$ are characterized parameters varying with diffident network topology, and $\upsilon, \psi, \xi$ and $\zeta$ are nonlinear control parameters. Based on the above definition, the communication impedance depends on the node degree, link distance, frequency of communication, average signal energy noise ratio and the cellular switching times. First, a vehicle with a large degree or high betweenness centrality means it participating in quantities of communication missions, which leads to a relatively long store-and-forward delay and high probability of blocking. Second, long communication distance conduces high path loss and consumes much more signal power. What is more, a small cellular radius leads to more cell handovers $n_{s}$, which also increases the time delay and deteriorates the communication performance. In these two aspects, the communication impedance should be positively correlated with $k$, $B$ and $n_{s}$. Third, a high average signal energy noise ratio per unit distance contributes a robust communication, naturally being negatively correlated to the impedance.
In this way, we have completely established a complex network graph model for the vehicular network communication.

\subsection{Complex Network Verification}
\label{3-3}
In this section, we quantitatively analyze and verify the small-world property and scaling-free property of the vehicular networks.
In the first place, we propose some key parameters depending on the complex network theory.

\emph{Node Degree Distribution}: The node degree of a vehicle $i$ in the vehicular network, marked as $k_{i}$, is defined as the number of the vehicles it can communicate with. Moreover, $p(k)$ is the probability that a randomized node's degree is $k$. And the distribution of $p(k)$ is defined as the node degree distribution.

\emph{Clustering Coefficients}: The characteristic that neighbors can also communicated with each other is called the clustering characteristic, which measures the tightness of the network. The vehicle $i$'s clustering coefficient is defined as the following:
\begin{equation}\label{16}
{{C}_{i}}=\frac{{{E}_{i}}}{{{k}_{i}}({{k}_{i}}-1)/2},
\end{equation}
where $k_{i}$ represents the node degree of vehicle $i$ and ${E}_{i}$ is the number of communication links among neighbors. Further more, the general clustering coefficient of the entire network is the average of ${C}_{i}$.

%\begin{figure}
%  \centering
%  \includegraphics[width=0.46\textwidth]{bc_model.eps}\\
%  \caption{Node importance analysis and judgement.}\label{bc_model}
%\end{figure}
\emph{Betweenness Centrality}: The normalized betweenness centrality $B$, and therefore, is defined to measure the importance of the node from another dimension, i.e.,
\begin{equation}\label{19}
{B_{i}}=\frac{2}{(N-1)(N-2)}\sum\limits_{s\ne i\ne t}{\frac{n_{st}^{i}}{{{g}_{st}}}},
\end{equation}
where ${{g}_{st}}$ is the number of the shortest path from $s$ to $t$, and $n_{st}^{i}$ notes the number of the shortest path via $i$ from $s$ to $t$.

\begin{figure}[!t]
  \begin{minipage}[t]{0.46\linewidth}
  \centering
  \includegraphics[width=1\textwidth]{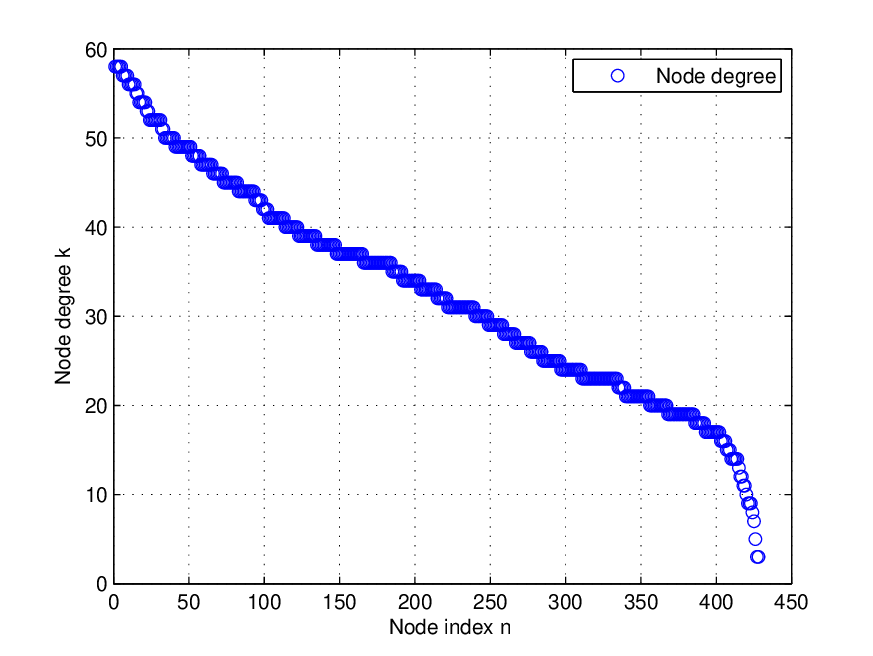}
  \centerline{\footnotesize{(a) Node Degree}}
  \end{minipage}
  \begin{minipage}[t]{0.46\linewidth}
  \centering
  \includegraphics[width=1\textwidth]{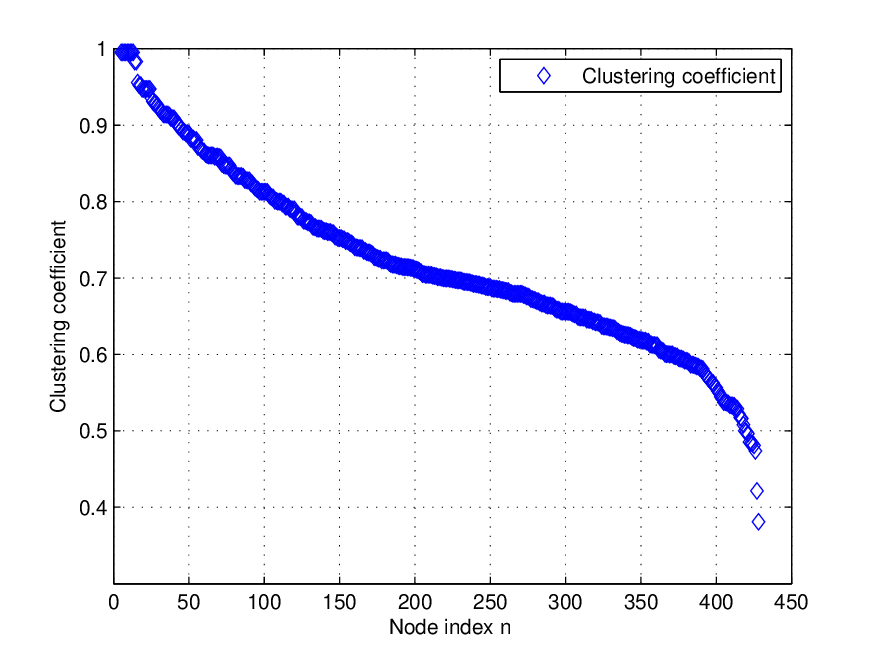}
  \centerline{\footnotesize{(b) Clustering Coefficient}}
  \end{minipage} \\
  \begin{minipage}[t]{0.46\linewidth}
  \centering
  \includegraphics[width=1\textwidth]{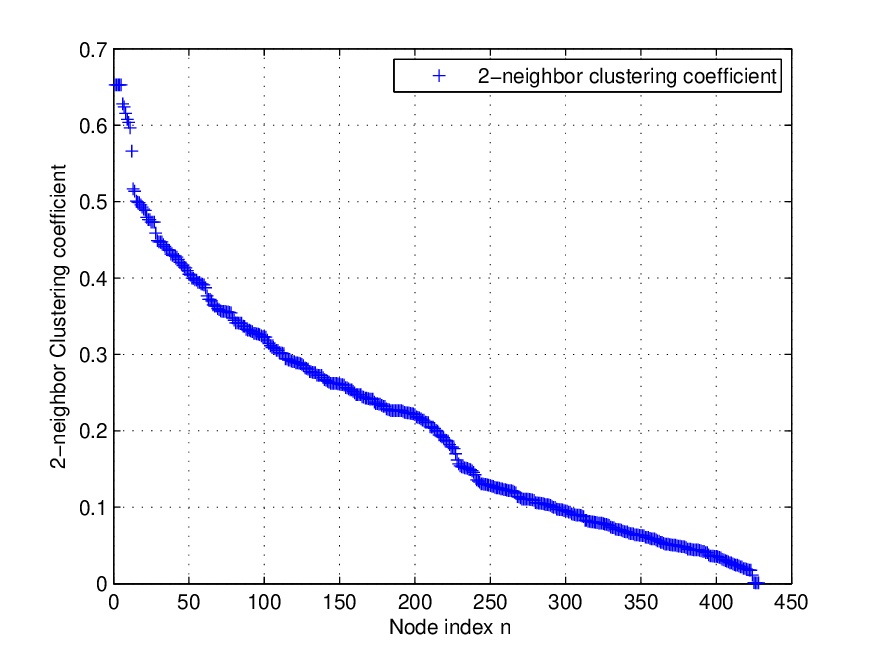}
  \centerline{\footnotesize{(c) 2-neighbor Clustering Coefficient}}
  \end{minipage}
  \begin{minipage}[t]{0.46\linewidth}
  \centering
  \includegraphics[width=1\textwidth]{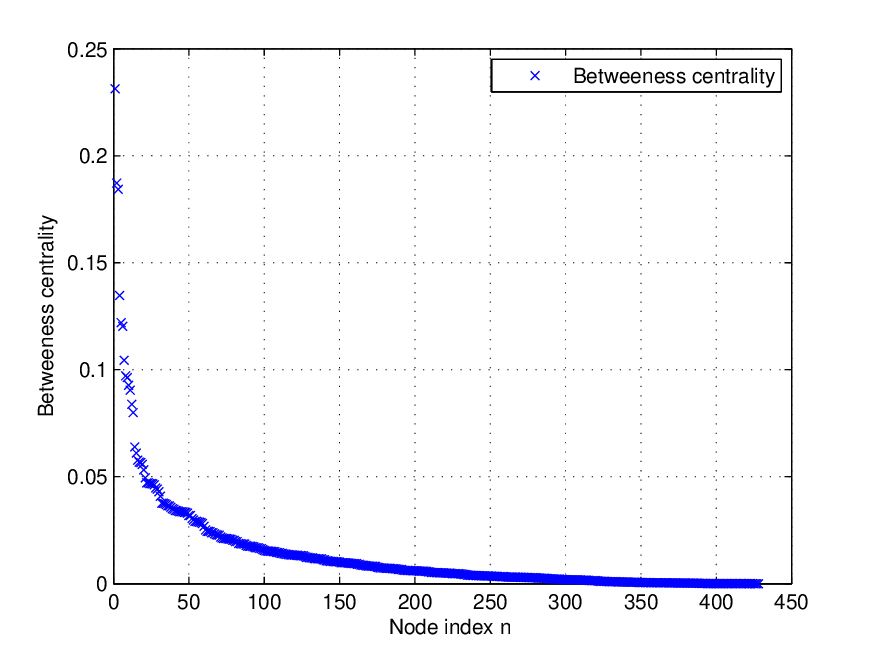}
  \centerline{\footnotesize{(d) Betweenness Centrality}}
  \end{minipage}
\caption{The Complex Network Parameters Verification.}\label{cc_bc}
\vspace{-4mm}
%\vspace{2.0cm}
\end{figure}

A data-driven numerical simulation is conducted for the vehicular network and we verify the complex network properties based on the Taxi GPS dataset. Fig.~\ref{cc_bc} demonstrates the parameters mentioned above of the proposed network with communication distance $r=500$. Moreover, we calculated the average network clustering coefficient $\overline{C}=0.7225$ and the average path length $\overline{l}=6.73374$.

The simulation results conform to the small world property (a high degree of clustering and a short average path length) and a scaling free distribution in node degree and betweenness centrality. In consequence, we can quantitatively treat the vehicular network as a complex network and the complex network theory bring us a new perspective in network design, optimization and management for the communication on vehicular networks. Next section, we will propose three optimization models under different communication models.

\section{Communication on the Vehicular Networks}
\label{The 5G Communication on the VANET}
 In Section~\ref{Data-Driven complex network model}, we have discussed the network topology of vehicular networks. Based on the analysis above, we establish the V2I (Section~\ref{Clustering Algorithm of the V2I Model}), V2V (Section~\ref{Traffic allocation on the V2V Model}) and the hybrid communication model (Section~\ref{Information Source Selection on the Hybrid Model}), respectively, with the communication impedance. Moreover, we propose a clustering algorithm for station selection, a traffic allocation optimization model and an information source selection model.
%\begin{figure}
%  \centering
%  \includegraphics[width=0.46\textwidth]{d2i.eps}\\
%  \caption{Device-to-infrastructure communication framework on the vehicular network.}\label{d2i}
%\end{figure}
\subsection{Clustering Algorithm of the V2I Model}
\label{Clustering Algorithm of the V2I Model}
In the following, we will focus on the V2I communication model. Similarly, the vehicle impedance in the V2I model is defined based on the Massive MIMO in vehicular communication system, which is a technology to enhance the overall networks performance. With a large excess of service antennas over terminals and time-division duplex operation, the extra antennas focuses energy into ever smaller regions of space and bring huge improvements in communication throughput and energy efficiency.
In~\cite{yang2014throughput}, the authors proposed the throughput $\widetilde{R_{k}}$ (achievable rate of the uplink transmission from user $k$ to measure the behavior of massive MIMO systems):
\begin{equation}\label{30}
\widetilde{{{R}_{k}}}\triangleq (1-\tau -\varsigma )\operatorname{E}[log(1+{{\gamma }_{k}})],
\end{equation}
where ${\gamma }_{k}$ shows the the signal-to-interference-plus-noise-ratio (SINR) which is a function containing channel model parameters and antennas parameters. $\tau$ is the channel estimation (CE) time, and $\varsigma$ is the wireless energy transfer (WET) time. In our model, we only consider the value of $\widetilde{{{R}_{k}}}$ instead of its impact factors. We assume that the base stations directly communicate with vehicles within its control range, which means that the distance from a vehicle to a base station is less than the cellular radius in the V2I Model. In this way, we define the communication impendence of vehicle $i$ as follows:
\begin{equation}\label{20}
{R_{i}}=\alpha {({{k}_{i}}{{B}_{i})}^{\upsilon}}+\beta {\widetilde{{{R}_{k}}}}^\psi, i=1,2,...,N.
\end{equation}
Similarly, $k_{i}$ represents the degree of the node $i$ and $B_{i}$ notes the betweenness centrality of the vehicle $i$. $\widetilde{{{R}_{k}}}$ shows the throughput of a certain vehicle-to-station communication link. Besides, $\alpha$ and $\beta$ are characterized parameters varying with diffident network topologies, while $\upsilon$ and $\psi$ are nonlinear control parameters.
A clustering algorithm based on the generalized distance $D$ is presented.
\begin{equation}\label{300}
{{D}_{ij}}=\epsilon ({{R}_{i}}+{{R}_{j}})+(1-\epsilon ){{d}_{ij}},
\end{equation}
where $R_{i}$ represents the vehicle impendence, $d_{ij}$ represents the realistic distance of two vehicles and $\epsilon$ denotes the weighting coefficient.

\noindent\rule[0.25\baselineskip]{0.49\textwidth}{1pt}

\noindent \textbf{Clustering algorithm based on generalized distance.}

\noindent Step1: Select one sample point as the clustering center $c_{1}$.

\noindent Step2: Calculate the generalized distances to the center, and select the $i$ with $\max_{i} D_{ic_{1}}$ as center $c_{2}$.

\noindent Step3: Calculate all the generalized distances to the two centers, and select the $j$ with $\max \{\min \{{{D}_{j{{c}_{1}}}},{{D}_{j{{c}_{2}}}}\}\}$ as center $c_{3}$, the rest can be done in the same manner.

\noindent Step4: Based on the nearest neighbouring rule classifying other samples.

\noindent\rule[0.25\baselineskip]{0.49\textwidth}{1pt}

\begin{figure}
  \centering
  % Requires \usepackage{graphicx}
  \includegraphics[width=0.46\textwidth]{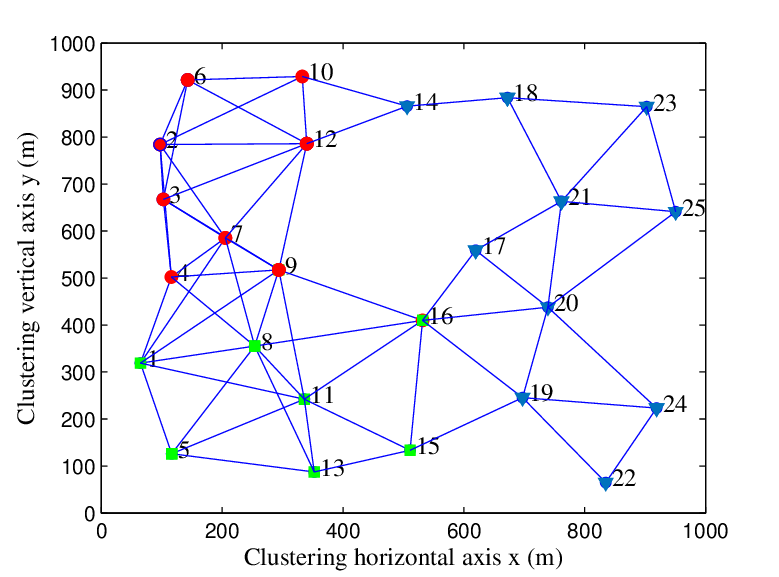}\\
  \caption{A clustering example based on the generalized distances (different color dots distinguishing the categories).}\label{clustering_ex}
  \vspace{-4mm}
\end{figure}

Fig.\ref{clustering_ex} shows a clustering example based on the generalized distance, which provides a constructive suggestion on the base station selection and cellular division.

\subsection{Traffic Allocation on the V2V Model}
\label{Traffic allocation on the V2V Model}
In terms of the complex communication missions in vehicular networks, a variety of services like real-time voice services, high definition video services and Internet access services should be supported whenever and wherever. However, how to allocate the communication traffic in an optimal fashion is worth discussing in details. For simplification, we assume that there are certain quantities of communication tasks transmitting from $n$ vehicles to a destination vehicle. The total communication demand quantity is marked as $Q$. Let $v$ be the vehicle node set and the starting vehicle set is denoted by $S={s_{1},s_{2},...,s_{n}}$ and $X={x_{1},x_{2},...,x_{n}}$ represents the allocated communication traffic allocation set, where $x_{i}$ is the actual communication task quantity on the $i$th communication link. We define the cost function $C(\mathbf{x})$ as:
\begin{equation}\label{199}
\begin{aligned}
   C(\mathbf{x})&=\sum\limits_{i=1}^{n}{\sum\limits_{u,v}{{{x}_{i}}R_{uv}^{i}}},
\end{aligned}
\end{equation}
where $R_{uv}^{i}$ is the communication impedance from vehicle $u$ to vehicle $v$ on the Dijkstra path under the condition of transferring the communication traffic $x_{i}$. Let $c$ be the communication capacity of each communication link, which denotes the maximum number of communication tasks and let $m_{uv}$ represents the total communication tasks on the communication link between vehicle $u$ and $v$, $m_{uv}\leq c$. We have the following optimization problem:
\begin{equation}\label{2001}
\begin{aligned}
 \min\ \ & C(\mathbf{x})=\sum\limits_{i=1}^{n}{\sum\limits_{u,v}{{{x}_{i}}R_{uv}^{i}}} \\
 s.t.\ \  & {{x}_{i}}\ge 0,\forall i=1,2,...,n, \\
     & \sum\limits_{i=1}^{n}{{{x}_{i}}\ge Q}, \\
     & {{m}_{uv}}=\sum\limits_{i=1}^{n}{{{x}_{i}}a_{uv}^{i}}\le c,\forall u,v\in V, \\
\end{aligned}
\end{equation}
where $\mathbf{x}={{[{{x}_{1}},{{x}_{2}},...,{{x}_{n}}]}^{T}}$ and $a_{uv}^{i}=1$, when the traffic $x_{i}$ goes through the link connecting the vehicle $u$ and $v$, otherwise $a_{uv}^{i}=0$. The network traffic allocation optimization problem can be casted as a convex optimization problem in (\ref{21}) by the definition of traffic-edge incidence matrix $\mathbf{A}\in \mathbf{R}^{E\times n}$, and
\begin{equation}\label{201}
{{\mathbf{A}}_{ij}}=\left\{ \begin{matrix}
   1, & \text{traffic }j\text{ passing the edge } i  \\
   0, & \text{otherwise}  \\
\end{matrix} \right.
\end{equation}
where $E$ is the total number of probable links, $\mathbf{x}={{[{{x}_{1}},{{x}_{2}},...,{{x}_{n}}]}^{T}}$, and $\mathbf{1}=[1,1,...,1]{{\text{ }}^{T}}$.
Then, we have
\begin{equation}\label{21}
\begin{aligned}
  \min\ \ & C(\mathbf{x}) \\
 s.t.\ \ & \mathbf{x}\ge \mathbf{0}, \\
 & {{\mathbf{x}}^{T}}\mathbf{1}\ge Q, \\
 & \mathbf{Ax}\le c\mathbf{1}. \\
\end{aligned}
\end{equation}
Furthermore, we can add a eigenfunction to this linear programming problem and rewrite it as follows:
\begin{equation}\label{24}
\begin{aligned}
\min\ \  &{{\mathbf{x}}^{T}}{{\mathbf{R}}_{w}}+\sum\limits_{i=1}^{n+E+1}{-\frac{1}{t}}\log (-f_{i}(\mathbf{x})) \\
s.t.\ \ & {{f}_{i}}(\mathbf{x})=-x_{i},i=1,2,...,n, \\
 & {{f}_{i}}(\mathbf{x})=Q-{{\mathbf{x}}^{T}}\mathbf{1},i=n+1, \\
 & {{f}_{i}}(\mathbf{x})=A_{i}\mathbf{x}-c,i=n+2,n+3,...,n+E+1. \\
\end{aligned}
\end{equation}
where $A_{i}$ represents the row vector of matrix $\mathbf{A}$ and auxiliary variable $t>0$ controls the computational accuracy. ${\mathbf{R}}_{w}$ is the sum of the communication impendence of each the allocation routing.

The solution of the problem (\ref{24}) is marked as $\mathbf{x^{\ast}}(t)$, which satisfies the condition:
\begin{equation}\label{29}
t{{\mathbf{R}}_{w}}-\frac{1}{\mathbf{x}}+\frac{1}{Q-{{\mathbf{x}}^{T}}\mathbf{1}}\cdot \mathbf{1}+{{\mathbf{A}}^{T}}\frac{1}{c\mathbf{1}-\mathbf{Ax}}=\mathbf{0},
\end{equation}
where let $\frac{1}{\mathbf{x}}\!=\!{{[\frac{1}{{{x}_{1}}},\frac{1}{{{x}_{2}}},...,\frac{1}{{{x}_{n}}}]}^{T}}$, $\forall\mathbf{x}\!\in\! {{\mathbf{R}}^{n}}$.
And we can prove that the deviation between $\mathbf{x^{\ast}}(t)$ and the optimal solution of primal problem is not more than $(n+E+1)/t$. Many computer simulation algorithms can solve the above optimization problem.

\subsection{Information Source Selection on the Hybrid Model}
\label{Information Source Selection on the Hybrid Model}
The criterion for selecting the information source location is to make the network capacity maximize. In another word, the information broadcasting facilities should be located near the source vehicles associated with information replicas. In this subsection, we focus on the hybrid communication model, where we study the optimal source vehicles selection strategy.
Let $q(i)$ indicate the probability of any packet to pass node $i$, and $n_{st}^{i}$ and ${g}_{st}$ are defined identically as (\ref{19}):
\begin{equation}\label{32}
q(i)=\sum\limits_{s(s\ne i)}{\sum\limits_{t(t\ne i)}{p(s,t)\frac{n_{st}^{i}}{{{g}_{st}}}}},
\end{equation}
where $p(s,t)$ is the probability of a packet to choose source vehicle $s$ and vehicle $t$ as its destination.
Instead of uniform distribution, the source vehicles obey the probability $p(s)$, while we assume that the destination vehicles of packets are uniformly distributed and are independently selected. We have:
\begin{equation}\label{34}
p(s,t)=p(s)p(t)=\frac{p(s)}{N-1}.
\end{equation}
Then, the probability of any packet to pass vehicle $i$ can be calculated as follows:
\begin{equation}\label{35}
 q(i)=\frac{1}{N-1}\sum\limits_{s\ne i}{\sum\limits_{t\ne i}{p(s)\frac{n_{st}^{i}}{{{g}_{st}}}}}.
\end{equation}
Define the $p(i|s)$ measuring the conditional probability of the situation where packet starts from vehicle $s$ to pass vehicle $i$,
\begin{equation}\label{36}
p(i|s)=\frac{1}{N-1}\sum\limits_{t(t\ne s, t\ne i)}{\frac{n_{st}^{i}}{{{g}_{st}}}}.
\end{equation}
Then, $R_{c}$ can be estimated as:
\begin{equation}\label{37}
  {{R}_{c}}=\frac{C}{{{\max }_{i}}\{{{R}_{i}}\sum\nolimits_{s}{p(s)p(i|s)}\}},
\end{equation}
where $R_{c}$ indicates the upper bound packets generated per time step to maintain in a flow state, and serves as a measure of the overall capacity of the network system, which is a function of betweenness centrality and communication impendence $R_{i}$.

%\begin{figure*}[!t]
%\begin{center}
%\subfigure[Degree distribution with maximum communication distance $r=100\mathrm{m}$]{\includegraphics[width=0.23\textwidth]{degree100.eps}}
%  \hspace{0.05in}
%\subfigure[Degree distribution with maximum communication distance $r=200\mathrm{m}$]{\includegraphics[width=0.23\textwidth]{degree200.eps}}
%  \hspace{0.05in}
%\subfigure[Degree distribution with maximum communication distance $r=500\mathrm{m}$]{\includegraphics[width=0.23\textwidth]{degree500.eps}}
%\hspace{0.05in}
%\subfigure[Degree distribution with maximum communication distance $r=800\mathrm{m}$]{\includegraphics[width=0.23\textwidth]{degree800.eps}}
%\caption{Degree distribution with different maximum communication distance.}
%\label{degree_communication}
%\end{center}
%\end{figure*}

The base station selection model, therefore, reduces to a a min-max problem:
\begin{equation}\label{39}
\begin{aligned}
\min \ \  & {{\max }_{i}}\{{{R}_{i}}\sum\nolimits_{s}{p(s)p(i|s)}\} \\
s.t. \ \ & 0\le p(s)\le 1, \\
 & \sum\limits_{s}{p(s)=1.} \\
\end{aligned}
\end{equation}
After introducing an auxiliary variable $\Lambda$:
\begin{equation}\label{40}
\Lambda ={{\max }_{i}}\{{{R}_{i}}\sum\nolimits_{s}{p(s)p(i|s)}\}(i=1,2,...,N),
\end{equation}
the optimization problem can be casted as a linear programming problem as follows:
\begin{equation}\label{41}
\begin{aligned}
\min \ \   & \Lambda  \\
s.t. \ \  & \mathbf{RAp}-\Lambda \mathbf{1}\le \mathbf{0}, \\
 & {{\mathbf{p}}^{T}}\mathbf{1}=1, \\
 & \mathbf{p}\ge \mathbf{0}, \\
\end{aligned}
\end{equation}
where $\mathbf{A}=[p(i|s)]$, $\mathbf{p}=[p(s),s=1,2,...,N]^{T}$ and $\mathbf{1}=[1,1,...,1]$.
$\mathbf{R}$ is defined in (\ref{411})
\begin{equation}\label{411}
\footnotesize
\mathbf{R}=\left[ \begin{array}{*{35}{l}}
   {{R}_{1}} & 0 & \cdots  & 0  \\
   0 & {{R}_{2}} & {} & \vdots   \\
   \vdots  & {} & \ddots  & 0  \\
   0 & \cdots  & 0 & {{R}_{N}}  \\
\end{array} \right].
\end{equation}

Thus, we can easily find the minimal $\Lambda$ by linear programming algorithms and get the numerical solution with the help of calculating computer.

\section{Simulation Results}
\label{Simulation Results}

In this section, we conduct simulation on the extensive studies about the network topology and the communication performances based on our models.
First of all, we analyze the influence of the maximum communication distance $r$ and other key communication parameters on the network topology.

Section~\ref{Traffic allocation on the V2V Model} proposed a vehicular network V2V communication model based on the complex network theory, relying on which we elaborated some complexity parameters to analyze the performance of the network in the respect of topology structure. In the following, we analyze the effect of communication parameters on the communication impedance. On this score, we only concentrate on the topology properties of the vehicular network based on the Taxis GPS in Beijing for the time being and give constructive suggestions on the traffic management and communication design.

\begin{figure*}[!t]
\begin{center}
\subfigure[The impact of carrier frequency on the impendence with different communication ranges.]{\includegraphics[width=0.23\textwidth]{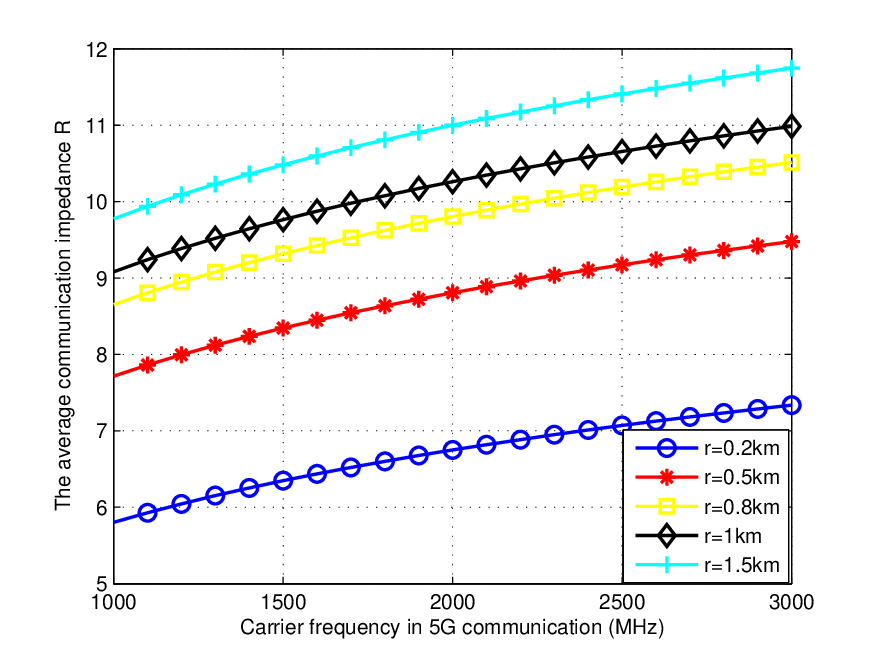}}
  \hspace{0.05in}
\subfigure[Cellular switching times with different communication ranges.]{\includegraphics[width=0.23\textwidth]{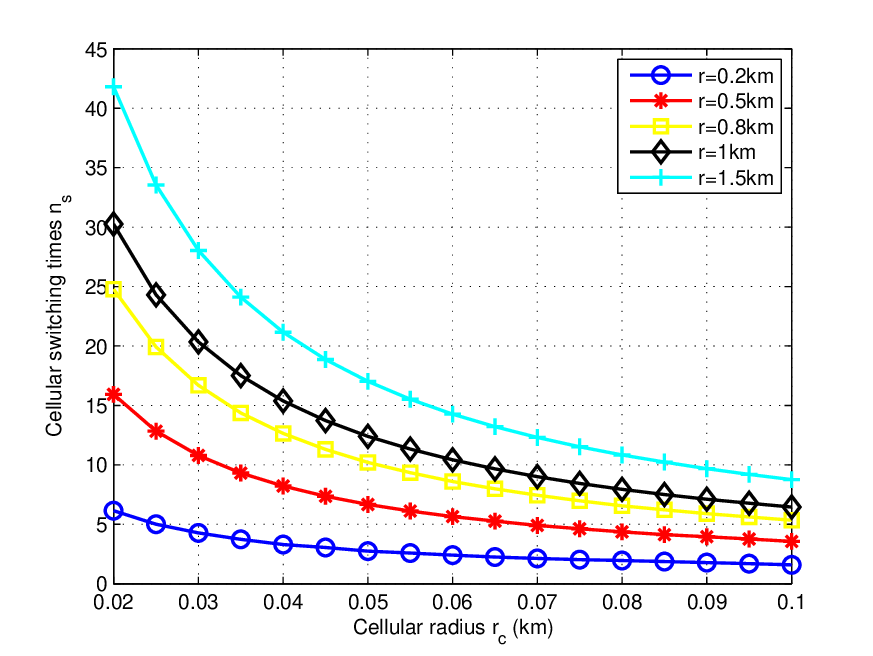}}
  \hspace{0.05in}
\subfigure[Communication impedance of each vehicle in the descend order.]{\includegraphics[width=0.23\textwidth]{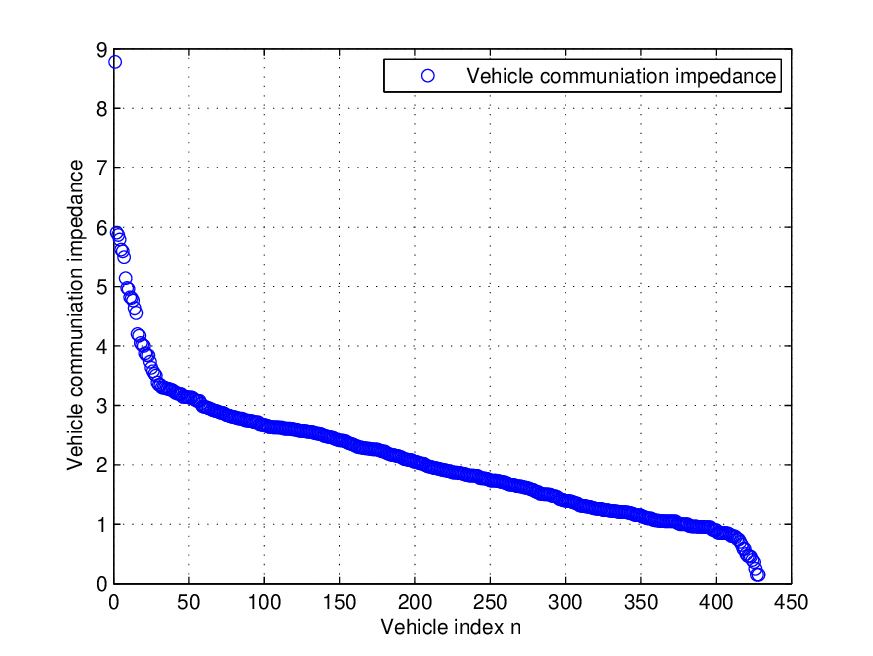}}
\hspace{0.05in}
\subfigure[Optimal source selection probability $p(s)$ distribution. ]{\includegraphics[width=0.23\textwidth]{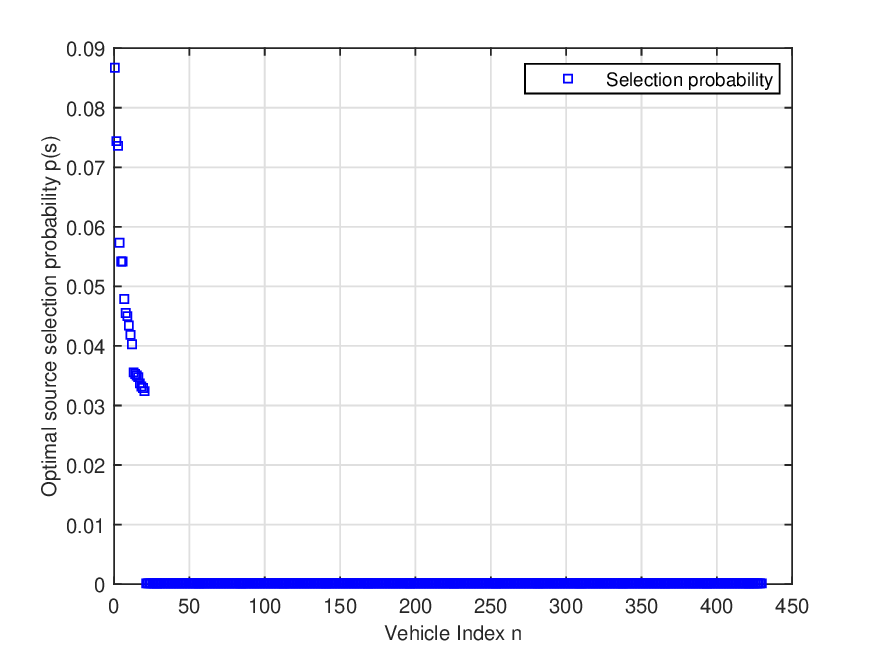}}
\caption{Communication impedances analysis and optimal information source selection on the hybrid model.}
\label{123}
\vspace{-7mm}
\end{center}
\end{figure*}

The carrier frequency mainly determines the transmission path loss $L_{u}$. We obtain five curves with different maximum communication distances, as in Fig.~\ref{123} subgraph (a). The vertical coordinates represents the average communication impedance for each of links and in this situation we neglect the effect of node importance by letting $\alpha=0$ in (\ref{15}). With the increasing of carrier frequency under each scenario, the average communication impedance $R$ ascends correspondingly. Obviously, the conclusion can be deduced from the definition of the communication impedance $R$. Likewise, a large maximum communication range $r$ contributes the communication impedance with more power loss. Specifically, with a small maximum communication range $r$ (200m$\sim$500m), communication impedance $R$ maintains a relatively small value but a high growth rate with $r$. However, when the $r$ reaches a specific distance (above 800m), the grow is slowing and communication impedance is tending towards stability. To our knowledge, the carrier frequency in communication may apply a high carrier frequency, but it needs a comprehensive consideration on the path loss and the communication range.
Fig. \ref{123} subgraph (b) shows the relationship between cellular radius and switching times under the condition of different maximum communication ranges. Generally speaking, the switching times descend with the increasing of cellular radius and the declining rate tends mildness. Even though we can improve spectrum efficiency and the transmission capacity by narrow down the cellular radius, large switching times reduce the communication performance in the same way. When we extend the maximum communication range, obviously there is a soaring incasement in the switching times under the scenarios of relatively small radiuses. As cellular radius $r_{c}>700m$, the switching times have no significant changes. The simulation results are consistent with the actual situation. The distribution of taxis in the city are concentrated in crowded areas, which is just the clustering feature of the small-world network. The average path length $l$ is surprisingly to a limited extent. As a consequence, in terms of an appropriate cellular radius $r_{c}$, the average switching times hover in a narrow range.
To summarize, the communication parameters to some extent affect the impedance of communication. In the realistic engineering, we should synthetically consider the carrier frequency, maximum communication distance, energy utilization efficiency, cellular radius etc, where a trade-off may contribute a communication effects. This paper provides a performance analysis method rather than the specific parameters.

As for the information selection model, Fig.~\ref{123} subgraphs (c) and (d) shows the related simulation results with the maximum communication distance $r=500\mathrm{m}$. Subgraph (c) demonstrates the communication impedance of each vehicle in the descend order. Subgraph (d) is the simulation result about how to select the information sources. Obviously, we can conclude that the vehicles play highly symmetrical roles in the information spreading. As shown in subgraph (d), only a few vehicles should act as sources in heterogeneous vehicular networks. That's means the source vehicle should be distributed within a small number of the nodes. Therefore, we can direct or manage fewer vehicles to control the entire vehicle network. More than that, an appropriate communication distance means a small range communication defined above due to the dispersed degree distribution and the low path loss. It makes great contribution to the green communication with a low power dissipation. As the vehicular network is a large-scale heterogeneous network, our work suggests that to improve the network capacity, information like traffic accident, congested roads or the traffic control should be broadcasted deriving from certain source vehicles.

\section{Conclusion}
\label{Conculsion}
In this paper, we analyzed the V2V and V2I communication performances on vehicular networks based on complex network theory. Furthermore, we proposed a clustering algorithm for station selection, a traffic allocation optimization model and an information source selection model, respectively which were viewed as examples for illustration of the concrete application of the defined communication impedance.

\section*{Acknowledgment}
This research was supported by the NSFC China under projects
61371079, 61271267 and 91338203.

% conference papers do not normally have an appendix

% use section* for acknowledgement
%\section*{Acknowledgment}
%
%This work was supported by a grant from the National Natural Science Foundation of China, No. 91338203, 61371079, 61271267, 60932005, and from the Ph.D. Programs Foundation of Ministry of Education of China, No. 20110002110060, and from the National Ocean Key Laboratory Open Foundation, No. 20131450166, and from the Tsinghua Sci-Tech Project, No. 2011THZ0.

% trigger a \newpage just before the given reference
% number - used to balance the columns on the last page
% adjust value as needed - may need to be readjusted if
% the document is modified later
%\IEEEtriggeratref{8}
% The "triggered" command can be changed if desired:
%\IEEEtriggercmd{\enlargethispage{-5in}}

% references section

% can use a bibliography generated by BibTeX as a .bbl file
% BibTeX documentation can be easily obtained at:
% http://www.ctan.org/tex-archive/biblio/bibtex/contrib/doc/
% The IEEEtran BibTeX style support page is at:
% http://www.michaelshell.org/tex/ieeetran/bibtex/
%\bibliographystyle{IEEEtran}
% argument is your BibTeX string definitions and bibliography database(s)
%\bibliography{IEEEabrv,../bib/paper}
%
% <OR> manually copy in the resultant .bbl file
% set second argument of \begin to the number of references
% (used to reserve space for the reference number labels box)
\bibliographystyle{IEEEtran}
\bibliography{ref}

% Generated by IEEEtran.bst, version: 1.13 (2008/09/30)
\begin{thebibliography}{10}
\providecommand{\url}[1]{#1}
\csname url@samestyle\endcsname
\providecommand{\newblock}{\relax}
\providecommand{\bibinfo}[2]{#2}
\providecommand{\BIBentrySTDinterwordspacing}{\spaceskip=0pt\relax}
\providecommand{\BIBentryALTinterwordstretchfactor}{4}
\providecommand{\BIBentryALTinterwordspacing}{\spaceskip=\fontdimen2\font plus
\BIBentryALTinterwordstretchfactor\fontdimen3\font minus
  \fontdimen4\font\relax}
\providecommand{\BIBforeignlanguage}[2]{{%
\expandafter\ifx\csname l@#1\endcsname\relax
\typeout{** WARNING: IEEEtran.bst: No hyphenation pattern has been}%
\typeout{** loaded for the language `#1'. Using the pattern for}%
\typeout{** the default language instead.}%
\else
\language=\csname l@#1\endcsname
\fi
#2}}
\providecommand{\BIBdecl}{\relax}
\BIBdecl

\bibitem{hartenstein2008tutorial}
H.~Hartenstein and K.~P. Laberteaux, ``A tutorial survey on vehicular ad hoc
  networks,'' \emph{Communications Magazine, IEEE}, vol.~46, no.~6, pp.
  164--171, June 2008.

\bibitem{giust2015distributed}
F.~Giust, L.~Cominardi, and C.~J. Bernardos, ``Distributed mobility management
  for future 5g networks: overview and analysis of existing approaches,''
  \emph{Communications Magazine, IEEE}, vol.~53, no.~1, pp. 142--149, Jan.
  2015.

\bibitem{demestichas20135g}
P.~Demestichas, A.~Georgakopoulos, D.~Karvounas, K.~Tsagkaris, V.~Stavroulaki,
  J.~Lu, C.~Xiong, and J.~Yao, ``5g on the horizon: key challenges for the
  radio-access network,'' \emph{Vehicular Technology Magazine, IEEE}, vol.~8,
  no.~3, pp. 47--53, Sept. 2013.

\bibitem{sharef2014vehicular}
B.~T. Sharef, R.~A. Alsaqour, and M.~Ismail, ``Vehicular communication ad hoc
  routing protocols: A survey,'' \emph{Journal of network and computer
  applications}, vol.~40, pp. 363--396, Apr. 2014.

\bibitem{zhang2015modeling}
H.~Zhang and J.~Li, ``Modeling and dynamical topology properties of vanet based
  on complex networks theory,'' \emph{AIP Advances}, vol.~5, no.~1, p. 017150,
  Jan. 2015.

\bibitem{jiang2014data}
C.~Jiang, Y.~Chen, and K.~R. Liu, ``Data-driven optimal throughput analysis for
  route selection in cognitive vehicular networks,'' \emph{Selected Areas in
  Communications, IEEE Journal on}, vol.~32, no.~11, pp. 2149--2162, Nov. 2014.

\bibitem{watts1998collective}
D.~J. Watts and S.~H. Strogatz, ``Collective dynamics of
  ¡®small-world¡¯networks,'' \emph{Nature}, vol. 393, no. 6684, pp. 440--442,
  June 1998.

\bibitem{barabasi1999emergence}
A.-L. Barab{\'a}si and R.~Albert, ``Emergence of scaling in random networks,''
  \emph{Science}, vol. 286, no. 5439, pp. 509--512, Oct. 1999.

\bibitem{yuan2010t}
J.~Yuan, Y.~Zheng, C.~Zhang, W.~Xie, X.~Xie, G.~Sun, and Y.~Huang, ``T-drive:
  driving directions based on taxi trajectories,'' in \emph{Proceedings of the
  18th SIGSPATIAL International conference on advances in geographic
  information systems}.\hskip 1em plus 0.5em minus 0.4em\relax New York: ACM,
  Nov. 2010, pp. 99--108.

\bibitem{correia2009view}
L.~M. Correia, ``A view of the cost 231-bertoni-ikegami model,'' in
  \emph{Antennas and Propagation, 2009. EuCAP 2009. 3rd European Conference
  on}.\hskip 1em plus 0.5em minus 0.4em\relax Berlin, Germany: IEEE, Mar. 2009,
  pp. 1681--1685.

\bibitem{yang2014throughput}
G.-M. Yang, C.-C. Ho, R.~Zhang, and Y.~Guan, ``Throughput optimization for
  massive mimo systems powered by wireless energy transfer,'' \emph{Selected
  Areas in Communications, IEEE Journal on}, vol.~30, no.~60, pp. 1--12, Jan.

\end{thebibliography}

% that's all folks
\end{document}